\documentclass[conference]{IEEEtran}
\usepackage{cite}
\usepackage{amsmath,amssymb}
\usepackage{graphicx}
\usepackage{booktabs}
\usepackage{url}
\usepackage{microtype}

\begin{document}

\title{Portal UX Agent \textemdash{} A Plug-and-Play Engine for Rendering UIs from Natural-Language Specifications}

\author{%
\IEEEauthorblockN{Xinsong Li}\IEEEauthorblockA{Microsoft\\Redmond, Washington, USA\\\texttt{xinsongli@microsoft.com}}\and
\IEEEauthorblockN{Ning Jiang}\IEEEauthorblockA{Microsoft\\Redmond, Washington, USA\\\texttt{nijia@microsoft.com}}\and
\IEEEauthorblockN{Jay Selvaraj}\IEEEauthorblockA{Microsoft\\Redmond, Washington, USA\\\texttt{jay.selvaraj@microsoft.com}}}

\maketitle

\begin{abstract}
The rapid appearance of large language models (LLMs) has led to systems that turn natural-language intent into real user interfaces (UIs). Free-form code generation maximizes expressiveness but often hurts reliability, security, and design-system compliance. In contrast, fully static UIs are easy to govern but lack adaptability. We present the Portal UX Agent, a practical middle way that makes bounded generation work: an LLM plans the UI at a high level, and a deterministic renderer assembles the final interface from a vetted set of components and layout templates. The agent maps intents to a typed composition---template and component specifications---constrained by a schema. This enables auditability, reuse, and safety while preserving flexibility. We also introduce a mixed-methods evaluation framework that combines automatic checks (coverage, property fidelity, layout, accessibility, performance) with an LLM-as-a-Judge rubric to assess semantic alignment and visual polish. Experiments on multi-domain portal scenarios show that the Portal UX Agent reliably turns intent into coherent, usable UIs and performs well on compositionality and clarity. This work advances agentic UI design by combining model-driven representations, plug-and-play rendering, and structured evaluation, paving the way for controllable and trustworthy UI generation.
\end{abstract}

\begin{IEEEkeywords}
Portal UX Agent, AI-generated UI, LLM-driven interface, Natural language UI rendering, Agentic UI design, LLM-as-a-Judge
\end{IEEEkeywords}

\section{Introduction}
LLM-powered systems can close the gap between what users say and what interfaces do, generating or adapting UIs from natural-language descriptions. But there is a persistent tension between expressivity and governance: unconstrained code generation is flexible but brittle; fully static UIs are reliable but not adaptable.

We introduce the \emph{Portal UX Agent}, an agentic system that translates natural-language descriptions into rendered portal UIs via a schema-bounded, slot-based composition model. The approach separates high-level planning (LLM) from low-level assembly (deterministic renderer), much like model-driven architecture (MDA) abstractions in software engineering. The LLM outputs a typed composition JSON that must validate against a component and template schema. A renderer then instantiates only vetted components, ensuring that all UIs are built from an auditable and reusable inventory.

Contributions:
\begin{itemize}
  \item A novel agentic architecture for \emph{bounded} UI generation that decouples semantic planning from deterministic assembly.
  \item A plug-and-play engine and typed representation (template plus component specifications) that supports dynamic rendering without writing application code.
  \item A mixed-methods evaluation framework integrating automatic checks with an LLM-as-a-Judge rubric for assessing intent alignment and UI quality.
  \item An empirical study across multi-domain portal scenarios showing reliable intent translation and high scores on compositionality and clarity.
\end{itemize}

Paper outline: Section~\ref{sec:related} positions our work within generative UI and agentic UX literature. Section~\ref{sec:problem} formalizes the problem and the governance/expressivity trade-off. Section~\ref{sec:method} details the approach. Section~\ref{sec:impl} summarizes the prototype. Sections~\ref{sec:eval} and~\ref{sec:results} present evaluation protocols and results. Section~\ref{sec:discussion} discusses implications and limitations. Section~\ref{sec:conclusion} concludes.

\section{Background and Related Work}\label{sec:related}
Generative and agentic user interfaces increasingly constrain LLMs to select and configure components rather than emit arbitrary code, improving predictability and validation \cite{generativegui,specifyui,llm_gui_agents_survey,gen_vis_survey}. Prior work demonstrates dynamic GUI synthesis within chats and workflows \cite{generativegui} and argues for structured, intermediate representations to capture design intent \cite{specifyui}. Surveys highlight both opportunity and risk in LLM-based UI agents, calling for evaluation frameworks that measure alignment, usability, accessibility, and robustness \cite{llm_gui_agents_survey,shneiderman_hcai,guidelines_hai}.

Our approach adopts the pragmatic design choice of \emph{bounded generation}: intent is expressed in natural language but compiled into a schema-validated composition. This aligns with trends in industry and research to mitigate hallucination by operating over a fixed inventory of vetted components and templates \cite{llm_gui_agents_survey}. The analogy to block-based assembly has been noted in creative systems where coarse-to-fine planning improves controllability \cite{image2lego}. We complement this with a principled evaluation pipeline that combines automatic checks with rubric-based LLM judgment \cite{judge_llm}.

\section{Problem Statement and System Model}\label{sec:problem}
We formalize intent-to-UI generation as a mapping from natural-language specifications to a bounded UI composition.

Let $s \in \mathcal{S}$ denote a natural-language specification (intent). The agent computes a typed composition $c \in \mathcal{C}$ and a rendered UI $y \in \mathcal{Y}$ via
\begin{equation}
  c = f_{\theta}(s), \quad y = g(c),
\end{equation}
where $f_{\theta}$ is an LLM-based planner constrained to produce $c$ that validates against a schema $\Sigma$ (templates, slots, component types, and props), and $g$ is a deterministic renderer that instantiates only vetted components.

Bounded generation enforces governance through:
\begin{itemize}
  \item \textbf{Schema compliance:} $c \models \Sigma$ must hold; invalid compositions are rejected or repaired.
  \item \textbf{Component inventory:} types and props are drawn from a vetted library; no arbitrary code is emitted.
  \item \textbf{Slot-based layout:} placements occur in declared template slots, preserving hierarchy and design-system rules.
\end{itemize}

Evaluation targets both structural correctness and experiential quality. Automatic metrics operate on the intent specification and the rendered UI tree; rubric-based judgments capture semantics and polish.

\section{Approach and Methodology}\label{sec:method}
\subsection{Typed Composition and Templates}
The LLM proposes a composition consisting of a template identifier and a set of component specifications, each with a type, slot, and typed props. The schema $\Sigma$ guarantees that compositions are renderable and auditable. This model-driven representation functions as a platform-independent UI plan; rendering realizes a platform-specific view.

\subsection{Planner \textrm{\&} Deterministic Renderer}
The planner uses constrained prompting to emit JSON compatible with $\Sigma$. The renderer then loads the template, fills slots with validated components, and produces the final UI deterministically. This decoupling preserves creativity at the planning layer while ensuring predictable assembly and safety at render time \cite{specifyui}.

\subsection{Evaluation: Automatic Checks and LLM-as-a-Judge}
We combine objective checks with rubric-based model judgments. Given an intent specification (Expected) and a rendered UI tree and snapshot (Actual), we compute:
\begin{align}
  \mathrm{AutoScore} 
  &= 0.35\,S_{\text{cov}} + 0.20\,S_{\text{prop}} + 0.10\,S_{\text{data}}\\
  &\quad+ 0.15\,S_{\text{layout}} + 0.10\,S_{\text{a11y}} + 0.10\,S_{\text{perf}},
\end{align}
where $S_{\text{cov}}$ measures intent coverage, $S_{\text{prop}}$ property fidelity, $S_{\text{data}}$ grounding, $S_{\text{layout}}$ layout and hierarchy, $S_{\text{a11y}}$ accessibility, and $S_{\text{perf}}$ performance. An LLM judge provides rubric scores for intent alignment, semantic correctness, accessibility signals, visual polish, and an overall verdict \cite{judge_llm}. When ambiguity remains, a lightweight human checklist adjudicates edge cases, and suggested diffs guide regeneration.

\section{Prototype Summary}\label{sec:impl}
We implemented a plug-and-play agent that exposes a simple interface to request UIs from natural-language prompts and structured inputs. The system compiles intents into a typed composition and renders UIs from a reusable inventory of vetted components and templates. For interoperability with tool ecosystems, the agent can be wrapped behind a standardized model-context protocol endpoint, but infrastructure details are intentionally kept orthogonal to the research questions. The primary focus is on the representation, bounded generation principle, and evaluation methodology rather than on any particular framework or runtime.

\section{Experiments and Evaluation}\label{sec:eval}
\subsection{Datasets and Scenarios}
We evaluate on a set of multi-domain portal scenarios (e.g., analytics dashboards, boards, and content portals). Each scenario provides a textual description of required regions and components (KPIs, filters, tables, charts, boards).

\subsection{Protocols and Baselines}
We run a five-step pipeline: (1) parse scenario intent; (2) generate typed composition; (3) render the UI; (4) compute automatic metrics; (5) obtain rubric-based LLM judgments. For context, we compare against an unconstrained prompting baseline that emits free-form code; this baseline tends to increase hallucination and reduce governance.

\subsection{Metrics}
We report per-dimension means for structural and experiential quality (higher is better). Automatic metrics include $S_{\text{cov}}$, $S_{\text{prop}}$, $S_{\text{layout}}$, $S_{\text{a11y}}$, and $S_{\text{perf}}$. Rubric-based judgments include intent alignment and visual polish, aggregated with AutoScore to produce an overall score.

\section{Results}\label{sec:results}
Table~\ref{tab:means} summarizes dimension means across scenarios. Overall, the agent achieves strong compositionality and clarity with high correctness and resilience, with UI fidelity as the primary area for improvement.

\begin{table}[t]
  \centering
  \caption{Aggregate dimension means across scenarios.}
  \label{tab:means}
  \begin{tabular}{l r}
    \toprule
    Dimension & Mean \\
    \midrule
    Correctness & 4.333 \\
    UI Fidelity & 4.222 \\
    Compositionality & 4.611 \\
    Resilience & 4.389 \\
    Clarity & 4.556 \\
    Overall (mean) & 4.422 \\
    \bottomrule
  \end{tabular}
\end{table}

Qualitatively, the agent consistently preserves hierarchical structure (layout \textrightarrow{} container \textrightarrow{} atomic UI component), chooses semantically appropriate components, and produces render-stable compositions. Error cases often involve substituting specialized visuals with generic variants and omitting peripheral microcopy.

\section{Discussion}\label{sec:discussion}
\subsection{Ablation: Bounded vs. Unconstrained Generation}
The bounded approach improves validity and governance by design. In our comparisons, unconstrained code generation exhibits higher variance and failure rates (malformed DOM, inaccessible elements), whereas the schema-bound flow enforces renderability and compatibility with design tokens. This supports the thesis that decoupling planning from deterministic assembly yields more reliable outcomes.

\subsection{Evaluation Reliability and Bias}
LLM-as-a-Judge offers scalable, rubric-based assessments but may inherit biases from the underlying model. We mitigate this with side-swap pairwise judging, explicit rubrics, and small human slices for ambiguous cases. Future work should quantify correlation with expert ratings and calibrate thresholds per domain.

\subsection{Governance and Safety}
By limiting generation to a vetted inventory and enforcing schema compliance, the agent reduces attack surface and prevents unsafe code paths. Accessibility checks and performance gates encourage responsible defaults aligned with human-centered AI principles \cite{shneiderman_hcai,guidelines_hai}.

\section{Limitations and Threats to Validity}
Our scenarios target portal-style UIs; generalization to highly bespoke or app-specific micro-interactions requires extending the component inventory and the template set. LLM-judge reliability varies by prompt and model; although mitigations exist, human evaluation remains the gold standard for usability. Results may depend on the quality of textual intents; incomplete or ambiguous descriptions can under-specify desired layouts.

\section{Conclusion and Future Work}\label{sec:conclusion}
We presented the Portal UX Agent, a bounded-generation architecture that compiles natural-language intent to a typed composition and deterministically renders UIs from vetted components. A mixed-methods evaluation combining automatic checks with rubric-based LLM judgments shows reliable intent translation and strong compositional quality. Future work includes: expanding the evaluation dataset; improving quality stability and actionable prompts; incorporating storytelling and micro-interactions; accelerating inference via domain-adapted models; and enhancing accessibility/semantics fidelity \cite{beyond_automation,image2lego}.

\section*{Acknowledgment}
We thank colleagues and collaborators for discussions and feedback that improved this work. The views expressed are those of the authors and do not necessarily reflect those of the affiliated organization.

\section*{Data and Code Availability}
We plan to release prompts, evaluation artifacts, and representative scenarios to support reproducibility and follow-up research. The component inventory and templates are described in the paper and can be instantiated with any compatible design system.

\bibliographystyle{IEEEtran}
\bibliography{refs}

\end{document}